# BTAS: A Library for Tropical Algebra

Ahsan Humayun
Research Scholar
Department of Computer Science
National Textile University,
Faisalabad Pakistan
ahsanhumayun.ah@gmail.com

Dr. Muhammad Asif
Assistant Professor
Department of Computer Science
National Textile University,
Faisalabad Pakistan
asi135@gmail.com

Dr. Muhammad Kashif Hanif
Assistant Professor
Department of Computer Science
Government College University,
Faisalabad Pakistan
mkashifhanif@gmail.com

*Abstract—* **GPUs are dedicated processors used for complex calculations and simulations and they can be effectively used for tropical algebra computations. Tropical algebra is based on max-plus algebra and min-plus algebra. In this paper we proposed and designed a library based on Tropical Algebra which is used to provide standard vector and matrix operations namely Basic Tropical Algebra Subroutines (BTAS). The testing of BTAS library is conducted by implementing the sequential version of Floyd Warshall Algorithm on CPU and furthermore parallel version on GPU. The developed library for tropical algebra delivered extensively better results on a less expensive GPU as compared to the same on CPU.**

## I. INTRODUCTION

Tropical algebra is a relatively new mathematics field. The term tropical was first used by French mathematicians [1] in respect of Brazilian friend Imre Simon [2] who is considered as pioneer of tropical algebra. Initially tropical algebra was only developed within the context of discrete mathematics and optimization but later on scientists realized the power of its applications in different fields like computational algebra [3], discrete event simulation [4], operating systems [5], dynamic programming [6], modeling communication networks [7], biological sequence comparisons, social network analysis, petri net, hidden Markov models [8], and other related applications.

Tropical algebra is a discrete algebraic system to represent and solve real world problems [9]. Tropical algebra is based on max-plus algebra and min-plus algebra. In max-plus algebra, the tropical sum of two numbers is their maximum and the tropical product of two numbers is their sum [10]. Equations (1) and (2) depict these operations.

$$x \oplus y = \max(x, y) \quad (1)$$
$$x \odot y = x + y \quad (2)$$

Similarly, in min-plus algebra the tropical sum of two numbers returns the minimum among the two numbers while the tropical product of two numbers returns the sum of those two numbers, [10] as shown in equations (3) and (4).

$$x \oplus y = \min(x, y) \quad (3)$$
$$x \odot y = x + y \quad (4)$$

### Related Work

Tropical algebra was at its outset in the initial time period, yet it has since developed when the researchers realize the importance and practical application areas, and it has now become an essential part of combinatory and linear algebra. Basic Linear Algebra Subprograms (BLAS are routines that offer standardized ingredients for executing basic vector and matrix operations. CUDA Basic Linear Algebra Subprograms CUBLAS, It was developed by Nvidia's GPU/CUDA with contributions from Vasily Volkov, Davide Barbieri and from the University of Tennessee [11]. CUDA is a computing platform that is intended to give opportunity for the development of parallel operations [6]. Matrix Algebra on GPU and Multi-core Architectures (MAGMA) is a group of dense linear algebra routines, a successor of Lapack and ScaLapack, specially developed for heterogeneous GPU-based architecture.

## II. PROBLEM FORMULATION

Serial and parallel libraries for linear algebra such as BLAS do exist. This research focuses to provide a library to perform parallel vector and matrix operations using tropical algebra. Researchers from different scientific areas solve their problems using tropical algebra [12]. However, there is no standardized library for such operations.

## III. PROPOSED METHODOLOGY

The proposed library BTAS will focus on providing routines for basic tropical algebra operations. The contribution of BTAS is to facilitate the people from different research groups with advance, highly optimized outcome based tropical algebra library that is user friendly, extensible and freely accessible without any cost. It is important since it serves as dependence for researchers from different areas that have to develop code for problems which can be solved using tropical algebra. BLAS computations on general-purpose machines have been studied extensively [13]. Furthermore, in recent years, GPUs have become a popular target for acceleration. So it is feasible to use modern general-purpose GPUs (GP-GPUs) effectively for tropical algebra computations. In this paper, we implemented the solution which is based on matrix multiplication to address the all pairs shortest path problem on GPU. The given weighted graph is multiplied with itself about n-1 times and hence it provides the solution of APSP. On the other hand, repeating squaring strategy decreases the time complexity up to $O(n^3 \log_2(n))$ for APSP. The results will be compared on the basis of speedup and runtime between the execution on CPU and GPU. We have used the Amazon EC2 instance as an operating environment for the execution and evaluation. Amazon EC2 is a high performance public cloud computing service facilitated by Amazon Web Services (AWS) which provides the Cluster GPU Instances (CGI). The characteristics of the Amazon EC2 services are presented in (Table 1)

TABLE I. SPECIFICATION OF AMAZON EC2 GPU INSTANCE (CGI)

| | |
|---|---|
| **Central Processing Units** | 2 × Intel Xeon Quad core X5570 2.93 GHz (Each CPU gives 46.88 GFLOPS in Double Precision) |
| **Elastic Cloud 2 Computing Units** | 33.5 |
| **Graphic Processing Unit** | 2 × "Fermi" M2050 (515 GFLOPS DP each GPU) |
| **Random Access Memory** | 22 GB |
| **Hard Disk** | 1600 GB |
| **Virtual Server** | Xen Hyper Virtual Machine 64-bit |

## IV. SIMULATIONS AND RESULTS

The Floyd Warshall algorithm for computing the required time of sequential execution is implemented on the CPU. Single thread execution methodology is used for the implementation of the algorithm. The pseudo code of the algorithm is as follow:

**Algorithm 2- Floyd Warshall Execution on CPU**
```
1.  void FW (int a[][10])
2.  {
3.  int x,y,z;
4.  for (z = 0; z < 10; z++)
5.  {
6.  for (x = 0; x < 10; x++)
7.  {
8.  for (y = 0; y < 10; y++)
9.  {
10. if ((a[x][z] * a[z][y] != 0) && (x != y))
11. {
12. if ((a[x][z] + a[z][y] < a[x][y]) || (a[x][y] == 0))
13. {
14. a[x][y] = a[x][z] + a[z][y];
15. }}}}}
16. for (x = 0; x < 10; x++)
17. {
18. cout<<"n Least Calculated x<<endl;
19. for (y = 0; y < 10; y++)
20. {
21. cout<<a[x][y]<<"t";
22. }}}
```

The Floyd-Warshall algorithm applies the dynamic programming methodology and executes in O ($n^3$) time. The algorithm takes a coordinated graph G = (V,E) with non-negative weights on the edges. It is simple to implement on CPU, thus it demonstrates that the graph does not required to be copied and hence keeps the same memory amount.

The similar algorithm is executed on GPU for the correlation of its calculation time with the time it takes to execute on CPU and for measuring effectiveness and speedup. The bigger matrix is further more divided into small matrices and each matrix is allocated to single thread. All matrices are executed on individually assigned threads concurrently and hence a massive speed up and efficiency is achieved by GPU execution.

The pseudo code for allocation of matrix on the threads is given as follow:

**Algorithm 3- Floyd Warshall Execution on GPU**
```
1.  GPU_FW(int *matr_f, int w, int M)
2.  {
3.  int a = blockIdx.x * blockDim.x + threadIdx.x;
4.  int b = blockIdx.y * blockDim.y + threadIdx.y;
5.  if(a<M&&b<M){
6.  int a0 = a*M + b;
7.  int a1 = a*M + w;
8.  int a2 = w*M + b;
9.  if(matr_f[a1] != -1 && matr_f[a2] != -1)
10. matr_f[a0] =
11. (matr_f [a0] != -1 && matr_f [a0] < matr_f [a1] + matr_f [a2])
12. ? matr_f [a0] : (matr_f [a1] + matr_f [a2]);
13. }}
```

Code optimization is a troublesome undertaking with numerous disadvantages on GPU. The comparison shows the better performance results on a GPU rather than CPU. Through experimentation we have examined that parallelization of the code has a significant impact on the performance [14].

**Algorithm 3- Matrix-Matrix Multiplication for Floyd-Warshall Algorithm**
1. **Matrix multiplication & addition through linear algebra**
2. Z ← Z + X × Y
3. z (i , j ) ← z(i , j ) + $\sum_{k=1}^{n} x(i,k) \cdot y(k,j)$
4. **Matrix multiplication & addition through min-plus algebra**
5. **Replace + by min and × by +.**
6. Z ← Z ⊕ X ⊙ Y
7. z(i , j ) ← min(z (i,j) $\min_{k=1}^{n}$(x(i,k) + y(k,j)))

The given input in the form of matrix is stored by using the Adjacency matrix, after that the kernel of the GPU is launched. As the GPU kernel starts its execution the bigger matrix is divided into smaller matrix and individual threads are assigned to each matrix. Each of thread completes its execution and returns the results to the kernel of GPU. The kernel compiles the execution of all the threads and sends it to CPU, and hence the CPU presents the results as an output.

The block diagram of implementing the APSP problem is shown in (Figure 1).

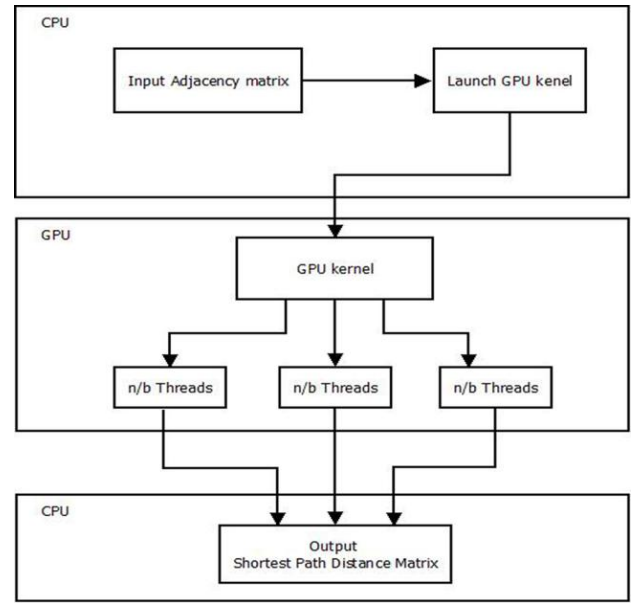

**Figure 1: Block Diagram of APSP Execution on GPU**

The implementation on GPU produce a significant results that shows the high increase in performance over the CPU implementation, and gives an average performance increase of over Floyd-Warshall algorithm testing. In this research, High level of data parallelism is achieved by utilizing massive computational power of GPUs to upgrade the algorithms based upon dynamic programming. The execution of the algorithms that are appropriate or supported for the vector-processing architecture of GPU can improve the performance of dynamic programming algorithms.

In serial execution, the Floyd-Warshall algorithm is a renowned solution that is given by the factors i.e. time-complexity $O(V^3)$ and space complexity $O(V^2)$. In parallel execution, the CUDA kernel code implements from line 4 of Algorithm 3. This will require just the last output size to be of $O(V^2)$. All of the intermediary executions do not require this memory. Hence the final output can be saved in the CPU memory.

CPU takes around 0.008ms in order to complete its execution on the given algorithm whereas the GPU takes around 0.001ms for the completion of its execution. The comparison of both executions is shown in (Figure 2).

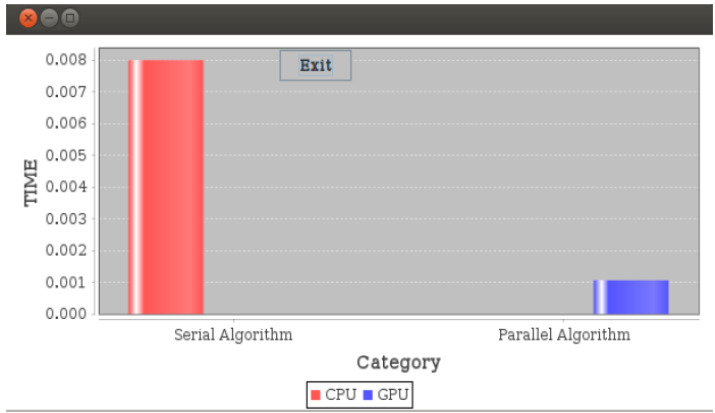

**Figure 2 Comparison of Execution on CPU and GPU**

We have investigated various data sets for Floyd-Warshall Algorithm, Moreover to confirm our execution performance more precisely, Weproduced random matrix by expanding number of vertices that reaches out from 4 to 128. The processing time taken by GPU and CPU to execute matrices for different data sets is figured. The computational accelerate achieved by the GPU over CPU is observed and shown in (Table 2).

TABLE II. COMPARISON OF EXECUTION ON CPU AND GPU

| Matrix | CPU Time | GPU Time | GPU Time Speed up |
|---|---|---|---|
| 4 | 0.008 | 0.001062 | 0.067 |
| 16 | 0.078 | 0.001163999 | 0.188 |
| 32 | 0.543 | 0.002068 | 0.677 |
| 64 | 4.034 | 0.0011743 | 2.274 |
| 128 | 31.066 | 0.0048330000 | 7.082 |

## V. RESULTS AND DISCUSSIONS

The transformation of dynamic programming algorithms into matrix-matrix product is quite tedious job because of dependencies in data elements. Moreover limitations like memory allocation and management should also be tackled in order to produce a matrix-matrix product based solution. We have developed parallel library for tropical algebra based problems using matrix multiplication. The solution based on matrix-matrix product gives better performance and speed-up as compared to dynamic programming approach and achieve great latency, throughput and speed difference as compared to serial implementation. The Floyd-Warshall algorithm is a member of APSP problem. We have implemented the matrix-matrix product solution on GPU and tested it against the Floyd-Warshall algorithm. The results depicts speed-up of for several data sets considered. In this research, we have developed the library for tropical algebra which delivered extensively better results on a less expensive GPU as opposed to on the CPU though numerous issues can't be determined through GPU programming since they can't be executed through matrix-matrix multiplication. The inapplicability of implementation of these problems is because of latency issues, communicating data from CPU to GPU and memory issues those necessities much GPU-CPU correspondence.